# Optimal control driven functional electrical stimulation: A scoping review

*Kevin Co[a,*], Mickaël Begon[a,b], François Bailly[c], Florent Moissenet[a,d,e]*

[a]*Laboratoire de Simulation et Modélisation du Mouvement, Université de Montréal, Montreal (Qc), Canada*
[b]*Research center Azrieli of the CHU Sainte-Justine, Montreal (Qc), Canada*
[c]*CAMIN, Inria, Univ Montpellier, Montpellier, France*
[d]*Biomechanics Laboratory, Geneva University Hospitals and University of Geneva, Geneva, Switzerland*
[e]*Kinesiology Laboratory, Geneva University Hospitals and University of Geneva, Geneva, Switzerland*

## Abstract

**Introduction**: Rehabilitation after a neurological impairment can be supported by functional electrical stimulation (FES). However, FES is limited by early muscle fatigue, slowing down the recovery progress. The use of optimal control to reduce overstimulation and improve motion precision is gaining interest. This review aims to map the current literature state meanwhile clarifying the best practices, identifying persistent challenges, and outlining directions for future research. **Methods**: Following the PRISMA guidelines, a search was conducted up to February 2024 using the combined keywords "FES", "optimal control" or "fatigue" across five databases (Medline, Embase, CINAHL Complete, Web of Science, and ProQuest Dissertations & Theses Citation Index). Inclusion criteria included the use of optimal control with FES for healthy individuals and those with neuromuscular disorders. **Results**: Among the 44 included studies, half were *in silico* and half *in vivo*, involving 87 participants, predominantly healthy young men. Twelve different motor tasks were investigated, with a focus on single-joint lower-limb movements. These studies principally used simple FES models, modulating pulse width or intensity to track joint-angle. **Conclusions**: Optimal control-driven FES can deliver precise motions and reduce fatigue. Yet clinical adoption is slowed down by the lack of consensus about modelling, inconvenient model identification protocol and limited validation. Additional barriers include insufficient open-science practices, computational performance reporting and the availability of customizable commercial hardware. Comparative FES model studies and longitudinal trials with large cohorts, among other efforts, are required to improve the technology readiness level. Such advances would help clinical adoption and improve patient outcomes.

---

[*]*Corresponding author at Laboratoire de Simulation et Modélisation du Mouvement, Université de Montréal, Montreal (Qc), Canada*
*E-mail address: kevin.co@umontreal.ca (Kevin Co)*

## 1. Introduction

Neurological disorders such as stroke and spinal cord injury (SCI) impose a persistent societal and economic burden due to their high prevalence, long-term sequelae, and substantial impact on quality of life. The resulting functional impairments contribute to significant disability-adjusted life years [1] demanding efficient and personalized rehabilitation to optimize motor recovery. Among available approaches for restoring motor control, functional electrical stimulation (FES), which applies electrical pulses to excitable tissues to trigger action potentials in motor and sensory fibers, has shown considerable rehabilitation promises through neuroplastic changes [2–4]. Effective FES-based therapy relies on repetitive, functionally relevant movements known to enhance neuroplasticity [5], and on sustained patient engagement, which correlates with improvements in voluntary muscle activation, cardiovascular health, and reductions in muscle atrophy and spasticity [6,7].

Despite its potential, FES faces some limitations. One of the primary challenges is the management of premature muscle fatigue, caused by the non-physiological activation of slow- and fast-twitch [8]. This rapid fatigue onset limits both the duration and intensity of rehabilitation sessions, slowing patient progress. Researchers have tested various countermeasures like tuning stimulation parameters such as frequency, pulse width and intensity [9,10], distributing the stimulation spatially and sequentially on the muscle [11,12], or combining FES with other external assistive devices during motor tasks [13]. Yet, due to the complex multiscale interactions involved in FES-evoked movement production, these solutions are rather empirical and rarely optimized for specific activities or tailored to individual patient characteristics. Accurately predicting fatigue, modeling the muscle's nonlinear behavior, and accommodating patient-specific range-of-motion constraints (from atrophy or spasticity) remain key challenges to developing truly personalized, fatigue-mitigating FES protocols.

Biomechanical models can help overcome these challenges by explicitly representing how FES, musculotendon units, bones, and joints interact during movement. By integrating patient-specific characteristics, such as range of motion, limb length, and nonlinear muscle properties (force-length-velocity relationships), they yield precise insights into each muscle's contribution [14]. When combined with FES models that map stimulation parameters to muscle dynamics, this framework enables personalized predictions of muscle force output and fatigue onset [15]With biomechanical representations, issues like muscle redundancy (multiple muscles driving the same degree of freedom), muscle synergy identification, and compensatory movement detection are significantly easier to manage. Unlike empirical approaches that heavily rely on trial and error, model-based control leverages comprehensive knowledge embedded within computational models to systematically address the complex multiscale interactions in FES-evoked movement production. This shift should not only avoid the inefficiencies of trial and error but also enable more reliable and efficient FES-based therapies.

This scoping review focuses on numerical optimal control methods, currently the most widely used forward-dynamics optimization approach for embedding patient-specific models in FES research. Optimal control is a mathematical framework that computes state variables and control inputs minimizing a specified cost function while satisfying the system's dynamics (e.g., muscle contraction and/or equations of motion typically described by ordinary differential equations) and any relevant constraints. Hence, by incorporating biomechanical models, it can notably enable the design of personalized stimulation patterns that can produce accurate, effective movements. In this context, a major strength of optimal control lies in its flexibility in optimizing stimulation parameters, thereby enhancing FES efficiency. Consequently, these optimized stimulation protocols hold promises for reducing FES-induced muscle fatigue and improving rehabilitation outcomes [16].

The earliest reported application of optimal control in FES dates back to 1997 [17], where a simple inverted-pendulum model was employed without accounting for muscle fatigue. Contemporary implementations, by contrast, require detailed musculoskeletal and FES models, patient-specific parameters for accurate motion, and real-time closed-loop optimization to handle factors like muscle fatigue. These enhancements, however, are computationally expensive, require sophisticated models and complex parameter identification procedure. Despite these challenges, optimal control-driven FES remains a promising strategy for coordinating multi-joint movements and tailoring stimulation to individual needs. Its effectiveness depends on a thorough understanding of FES and kinematic dynamic behavior and a carefully posed control problem, including the selection of appropriate biomechanical and FES models, and cost functions. Although many studies have tested optimal control in FES, critical questions remain: can it be applied in rehabilitation and can it reliably reduce muscle fatigue? This scoping review brings together the current evidence on optimal control in FES, aiming to clarify best practices, identify persistent challenges, and outline directions for future research.

## 2. Methods

This scoping review was conducted following the PRISMA-ScR guidelines [18] and the Joanna Briggs Institute framework [19]. The Population–Concept–Context approach was employed, including studies on healthy participants and those with neuromuscular disorders (Population), numerical optimal-control methods for personalizing and optimizing FES parameters (Concept), and studies conducted through simulations, controlled experiments, and clinical trials (Context).

### 2.1 Literature search

The search strategy was developed and executed with the support of a professional university librarian. We retrieved English-language articles, conference proceedings, and theses published up to February 9, 2024, from Medline, Embase, CINAHL Complete, Web of Science, and ProQuest Dissertations & Theses Citation Index. Search terms combined "functional electrical stimulation" with "optimal control" or "fatigue" (see supplementary File 1 for full search algorithms). Screening and data extraction were performed using Covidence [20].

### 2.2 Study screening

Two authors (KC, MB) independently screened records in two phases, first by title and abstract, then by full text, with disagreements resolved by a third author (FM). Studies involving animal models or examining muscle dynamics without explicit links to FES parameters were excluded [21,22]. A complementary manual cross-referencing search was also conducted to capture any additional relevant studies. Because this scoping review aims to provide an overview of the existing literature rather than generate clinical recommendations, no formal risk of bias assessment was conducted.

### 2.3 Data extraction

The data extraction form was first drafted using a representative study sample, then refined to cover the full range of included research. Data extraction was carried out by one author (KC) for consistency and independently validated by three others (FM, FB, MB) to ensure accuracy. This form included: **a)** General study details (i.e., publication year, format, journal, authors, study design, geographic origin); **b)** Objectives and hypotheses to characterize each work's focus; **c)** Participant/population data (i.e., sex, age, diagnosis, sample size); **d)** Intervention specifics (i.e., targeted muscles, functional tasks, FES devices, auxiliary equipment); **e)** Modeling parameters (musculoskeletal model: dimensions, degrees of freedom, muscle–tendon relationships; FES model: dynamic assumptions, rationale, optimized parameters and their ranges, and any fixed controls); **f)** Optimal-control problem formulation (i.e., cost function, discretization scheme, solver, control type (open- vs. closed-loop), computation time, hardware); **g)** Contextual notes (i.e., stated rationale for optimal control, reported challenges, and proposed future directions); **h)** Muscle-fatigue handling (if addressed, whether fatigue was mentioned, compensated, modeled, incorporated in the cost, tied to an FES parameter, or directly measured). When information was not explicit, details were traced through cited sources or inferred control type from cost functions and measurement methods.

**2.4 Data analysis**

To compare optimization complexity across studies and anticipate future trends, we proposed custom logarithmic complexity scores, to increase differences between approach complexities. These scores quantify the musculoskeletal model, FES model, and optimal control problem (OCP) complexity based on the extracted parameters:

$$MSK_{complexity} = \sqrt{\log\left(n_{muscle} \cdot n_{muscle_{relationship}} \cdot n_{dof}\right)^{d}} ,$$

with $n_{muscle}$ the number of used muscles, $n_{muscle_{relationship}}$ the number of considered relationships (e.g., passive force, force-length-velocity), $n_{dof}$ the number of freedom degree, and $d$ the model spatial dimension (e.g., 2 for 2D or 3 for 3D).

$$FES_{complexity} = \sqrt{\log\left(n_{FES_{optimparam}} \cdot n_{FES_{output}}\right)^{m}} ,$$

with $n_{FES_{optimparam}}$ the number of simultaneously optimized FES parameters, $n_{FES_{output}}$ the number of state output given by the model, and $m = 2$ if a fatigue model is included ($m = 2$ otherwise).

$$OCP_{complexity} = \sqrt{\left(\log\left(n_{costfunction} \cdot (1 + n_{constraint})\right) + n_{external_{device}}\right)^{c}} ,$$

with $n_{costfunction}$ the number of cost functions used, $n_{constraint}$ the number of constraints satisfied, $n_{external_{device}}$ the number of supplementary assistive devices, and $c = 2$ if closed-loop control ($c = 1$ otherwise).

To quantify the inclusion of muscle fatigue in optimal control-driven FES research, five closed questions were answered: *(i)* Was muscle fatigue mentioned in the study? *(ii)* Was muscle fatigue compensated for? *(iii)* Was muscle fatigue modeled in the simulation? *(iv)* Was muscle fatigue state incorporated in the cost function? *(v)* Was muscle fatigue measured during *in vivo* trials? These questions led to different combinations (see supplementary Fig. 5) and the studies were categorized into different levels from 0 (no inclusion of muscle fatigue) to 8 (full inclusion of muscle fatigue).

To assess the maturity of optimal control-driven FES for healthcare purposes, the technology readiness level (TRL) scale was used, ranging from TRL 1 (basic principles observed) to TRL 9 (proven system in clinical practice) [23]. This framework allows for a structured evaluation of the field's progression, from theoretical development to clinical implementation, and helps to identify the current barriers to clinical deployment [24]. In the context of this review, TRL levels were interpreted as follows: TRL 1–3: theoretical developments and *in silico* simulations; TRL 4: proof of concept demonstrated in lab conditions with healthy participants; TRL 5: validation in relevant environments (e.g., with pathological participants performing functional tasks), though not yet tested in real clinical workflows. Levels TRL 6 and above involve clinical trials, longitudinal validation, or independent replication.

## 3. Results

### 3.1 General study details

The search strategy yielded 1368 articles across the five selected databases (Fig. 1). After removing duplicates, 904 titles and abstracts were screened, of which 159 advanced to full-text review. Following screening validation and cross-referencing, 44 studies met all inclusion criteria and were retained for this scoping review. A selection of extracted information from included studies is reported in Tables 1 (lower limb tasks) and Table 2 (upper limb tasks). The full data extraction is available in supplementary File 2.

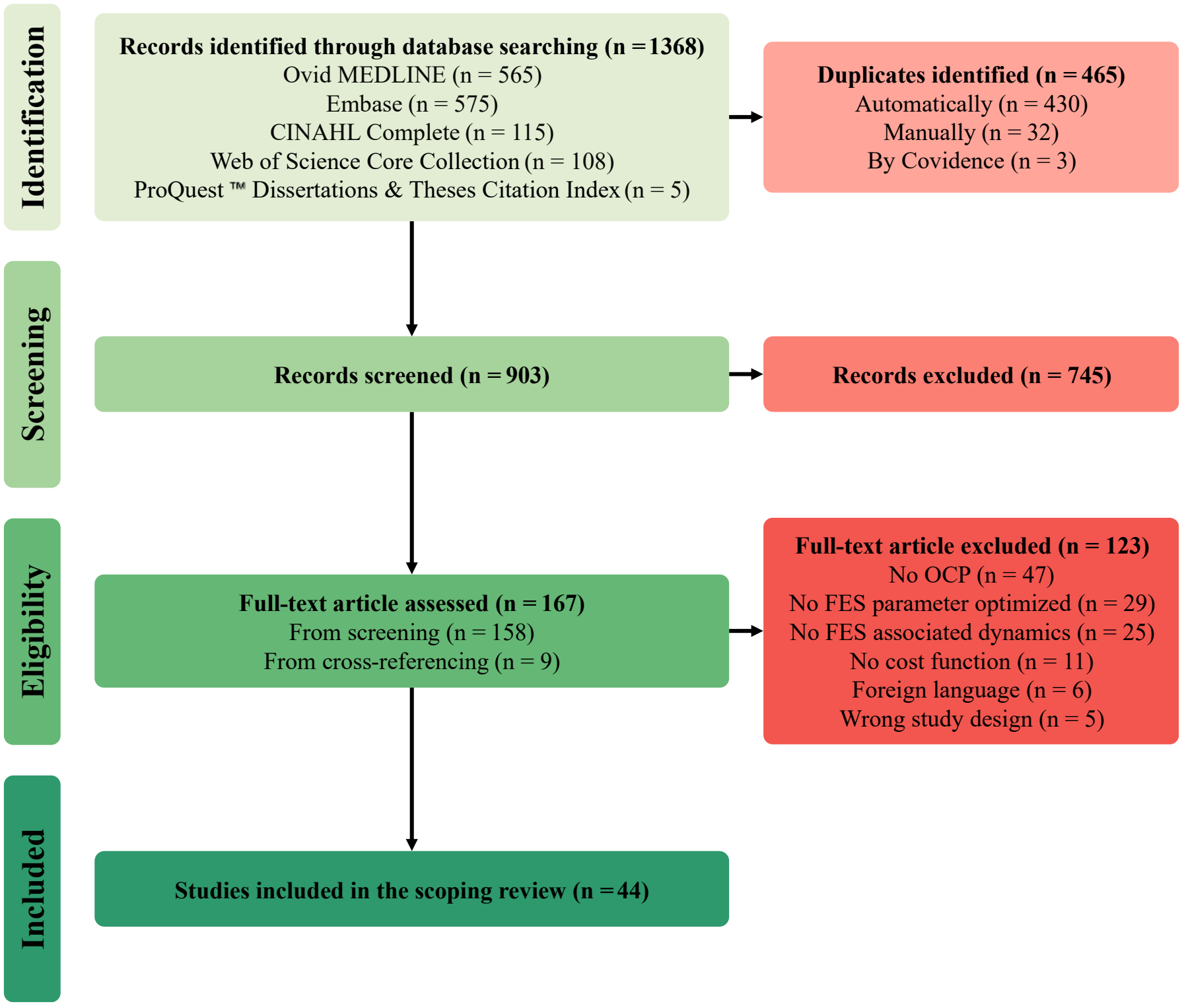

*Figure 1: PRISMA flow chart.*

Study designs were divided between *in silico* simulations (50%, n = 22 studies) and *in vivo* protocols (50%, n = 22 studies), encompassing 87 participants. The cohort was predominantly male (71%, p = 62 participants), healthy (68%, p = 59), and young (mean age 29 years [21.5–45.5]), with women representing 13% (p = 11) and 16% (p = 14) unreported. Among the 28 participants with neuromuscular disorders (32%), spinal cord injury was the most common (68%, p = 19), while stroke survivors accounted for 7% (p = 2) (see supplementary File 2 for full demographics). Researchers investigated 12 motor tasks, with a primary focus on lower limb movements (80%, n = 35 studies) over upper limb movements (20%, n = 9), knee extension being the most studied motor task (27%, n = 12). On average, two muscles, primarily mono-articular prime-movers, were stimulated. Stimulation parameters covered frequencies from 0 to 200 Hz (most common: 30 Hz) and pulse widths from 0 to 800 µs (most common:

400 μs). Pulse intensity was uniformly reported as a normalized value and varied depending on the stimulated muscle. Research targeting the lower limb most often applied 2D models with 2 degrees of freedom (DoF) or less (80%, n = 28 out of 35 studies) (Table 1). Concerning research targeting the upper limb, 2D models were used in 55% of them (n = 5 out of 9 studies) with only one study [25] employing a model with more than 2 DoF (Table 2). Overall, the primary aims were to develop (48%, n = 21) or refine (33%, n = 15) FES-based control systems, often coupled with motorized assistance (36%, n = 16), to enhance motor function (9%, n = 4), mitigate fatigue (18%, n = 8), or restore impaired movement (4%, n = 2). This research was performed by 14 different laboratories in 12 distinct countries (see supplementary Fig. 6 and 7).

*Table 1: Summary of studies* ***about lower limb movements*** *information (* optimized FES parameter).*

| **Studies** | **Study design** | **Partici-pants** | **Lower limb task** | **Muscle stimulated** | **Reported stimulation settings*** | **MSK model** *DoF \| Dim* | **FES model** | **Cost function** | **Constraints** | **TRL** |
|---|---|---|---|---|---|---|---|---|---|---|
| Bakir et al. (2020) [26] | *in silico* | N/A | Isometric contraction | N/I | Freq*: 0-T s<br>Width: N/A<br>Amp: N/A | None | Ding (2003) | *Min final force error* | *Dynamics AND FES bounds* | 3 |
| Bakir et al. (2022) [27] | *in silico* | N/A | Isometric contraction | N/I | Freq: 0-T s<br>Width*: 0-1 (-)<br>Amp: N/A | None | Modified Ding (2003) | *Min force error OR max muscle force* | *FES bounds* | 3 |
| Doll et al. (2015) [28] | *in vivo* | 1 AB | Isometric contraction | Quadriceps | Freq*: 0-200 Hz<br>Width: 900 µs<br>Amp: 35 mA | None | Ding (2003) | *Min force error AND min the number of delivered pulses* | *Dynamics AND FES bounds* | 4 |
| Doll et al. (2018) [29] | *in vivo* | 6 AB | Isometric contraction | Vastus medialis<br>Rectus femoris | Freq*: 15-40 Hz<br>Width: 900 µs<br>Amp: 40 mA | None | Ding (2003) | *Min force error AND min the number of delivered pulses* | *Dynamics AND FES bounds* | 4 |
| Zhang et al. (2013) [30] | *in vivo* | 2 AB | Isometric contraction | Tibialis anterior | Freq: 40 Hz<br>Width*: 0-450 µs<br>Amp: 26 or 35 mA | None | Zhang (2011) | *Min joint torque error AND min control variation* | *FES bounds AND Task duration bounds* | 4 |
| Bonnard and Rouot (2021) [31] | *in silico* | N/A | Isometric contraction | N/I | Freq*: 20 ms – T<br>Width: N/A<br>Amp: N/A | None | Ding (2003) | *Max final muscle force OR max final muscle force including final fatigue variable* | *Dynamics AND FES bounds* | 3 |
| | | | Knee extension | | | 1 \| 2 | Marion (2013) | | | |
| Bao et al. (2016) [32] | *in silico* | N/A | Knee extension | Quadriceps | Freq: N/A<br>Width: 140 µs<br>Amp*: 0-1 (-) | 1 \| 2 | Veltink (1992) AND Riener (1996) | *Min angle/velocity/activation error AND min stim amplitude AND fatigue AND device control* | *Dynamics AND Initial state AND Kinematic bounds AND Muscle activation bounds AND Fatigue state bounds* | 3 |
| Bao et al. (2019) [33] | *in silico* | N/A | Knee extension | Quadriceps | Freq: N/A<br>Width: N/A<br>Amp*: 0-1 (-) | 1 \| 2 | First order dynamic AND Riener (1996) | *Barrier function: min angle/activation/fatigue/motor* | *Dynamics AND Initial state AND FES bounds AND Fatigue state bounds* | 3 |

| Studies | Study design | Partici -pants | Lower limb task | Muscle stimulated | Reported stimulation settings* | MSK model *DoF \| Dim* | FES model | Cost function | Constraints | TRL |
|---|---|---|---|---|---|---|---|---|---|---|
| Sun et al. (2021) [34] | *in silico* | N/A | Knee extension | Quadriceps | Freq: N/A<br>Width: N/A<br>Amp*: Min/max threshold | 1 \| 2 | Veltink (1992) AND Riener (1996) | *Min joint angle / muscle activation error AND min muscle fatigue AND device control* | *Dynamics AND Terminal state AND FES bounds AND Muscle fatigue bounds AND Assistive device bounds* | 3 |
| Bao et al. (2021) [35] | *in vivo* | 1 AB<br>1 SCI | Knee extention | Quadriceps | Freq: 35 Hz<br>Width: 400 µs<br>Amp*: Min/max threshold | 1 \| 2 | Veltink (1992) AND Riener (1996) | *Min joint angle/activation error AND min muscle fatigue AND device control* | *Dynamics AND Initial state AND Terminal state AND Kinematic bounds AND FES bounds AND Assistive device bounds AND Disturbance bounds* | 5 |
| Benoussaad et al. (2015) [36] | *in vivo* | 4 SCI | Knee extension | Quadriceps | Freq: 20 Hz<br>Width*: 0-420 µs<br>Amp: level 3 (MRC) | 1 \| 2 | Makssoud (2011) | *Min joint angle error AND min muscle activation OR min knee power OR min muscle activation* | *Terminal state AND Kinematic bounds AND FES bounds* | 5 |
| Kirsch et al. (2017) [37] | *in vivo* | 3 AB | Knee extension | Quadriceps | Freq: 35 Hz<br>Width: 0-400 µs<br>Amp*: 0-1 (-) | 1 \| 2 | Veltink (1992) OR Schauer (2005) | *Min joint angle/pulse intensity/muscle activation error* | *Dynamics AND Initial state AND Muscle activation bounds AND FES bounds* | 4 |
| Kirsch et al. (2018) [38] | *in vivo* | 1 AB<br>1 SCI | Knee extension | Quadriceps | Freq: 35 Hz<br>Width: 400 µs<br>Amp*: 0-1 (-) | 1 \| 2 | Veltink (1992) AND Riener (1996) | *Min joint angle/joint velocity/muscle activation error AND min muscle fatigue AND motor control* | *Dynamics AND Initial state AND Muscle activation bounds AND FES bounds AND Assistive device bounds* | 5 |
| Sun et al. (2019) [39] | *in vivo* | 1 AB | Knee extension | Quadriceps | Freq: 35 Hz<br>Width: 400 µs<br>Amp*: 0-1 (-) | 1 \| 2 | Veltink (1992) | *Min joint angle/joint velocity/muscle activation error* | *Dynamics AND Initial state AND FES bounds* | 4 |

| Studies | Study design | Partici-pants | Lower limb task | Muscle stimulated | Reported stimulation settings* | MSK model *DoF \| Dim* | FES model | Cost function | Constraints | TRL |
|---|---|---|---|---|---|---|---|---|---|---|
| | | | | | | | | | *AND Task duration bounds* | |
| Benoussaad et al. (2007) [40] | *in silico* | N/A | Knee flexion-extension | Quadriceps<br>Hamstring | Freq: N/I<br>Width*: N/I<br>Amp*: N/I | 1 \| 2 | Makssoud (2004) | *Min joint angle error AND min muscle activation* | *FES bounds* | 3 |
| Benoussaad et al. (2008) [41] | *in silico* | N/A | Knee flexion-extension | Hamstring<br>Quadriceps | Freq: N/I<br>Width*: N/I<br>Amp: N/A | 1 \| 2 | Makssoud (2004) | *Min joint angle error AND min muscle activation OR min muscle activation* | *Terminal state AND Kinematic bounds AND FES bounds* | 3 |
| Wang et al. (2010) [42] | *in silico* | N/A | Knee flexion-extension | Quadriceps | Freq: N/A<br>Width: N/A<br>Amp*: 24-25.5 V | 1 \| 2 | Sharma (2009) | *Min joint angle error AND min FES activation* | *Dynamics AND Kinematic bounds AND Disturbance bounds* | 3 |
| Hunt et al. (1997) [17] | *in silico* | N/A | Standing | Ankle plantarflexor | Freq: 20 Hz<br>Width*: 0-500 µs<br>Amp: N/A | 1 \| 2 | Hunt (1998) | *Min joint angle AND joint torque AND FES activation* | *N/I* | 1 |
| Hunt et al. (1998) [43] | *in silico* | N/A | Standing | Gastrocnemius<br>Soleus | Freq: 20 Hz<br>Width*: 0-500 µs<br>Amp: N/A | 1 \| 2 | Hunt (1998) | *Min joint angle AND joint torque AND FES activation* | *N/I* | 3 |
| Holderbaum et al. (2002a) [44] | *in vivo* | 1 SCI | Standing | Lateral gastrocnemius<br>Soleus | Freq: 20 Hz<br>Width*: 0-500 µs<br>Amp: Muscle saturation | 1 \| 2 | Hunt (1998) | *Min joint angle error AND min disturbance influence* | *Mixed-sensitivity H∞ constraints* | 5 |
| Holderbaum et al. (2002b) [45] | *in vivo* | 1 SCI | Standing | Soleus<br>Lateral gastrocnemius | Freq: 20 Hz<br>Width*: 0-800 µs<br>Amp: Muscle saturation | 2 \| 2 | Hunt (1998) | *Min joint angle error AND min disturbance influence* | *Mixed-sensitivity H∞ constraints* | 5 |
| Mihelj and Munih (2004) [46] | *in vivo* | 1 SCI | Standing under perturbation | Gastrocnemius<br>Soleus<br>Tibialis anterior | Freq: 20 Hz<br>Width*: 0-1 (-)<br>Amp: N/A | 2 \| 2 | Hunt (1998) | *Min CoP deviation AND ankle torque* | *Dynamics AND Kinematic bounds* | 5 |
| Munih et al. (1997) [47] | *in vivo* | 1 AB<br>1 SCI | Standing under perturbation | Gastrocnemius<br>Tibialis anterior | Freq: 20 Hz<br>Width*: 0-500 µs<br>Amp: N/A | 1 \| 2 | Hunt (1998) | *Min joint angle AND joint torque* | *N/I* | 5 |

| Studies | Study design | Partici -pants | Lower limb task | Muscle stimulated | Reported stimulation settings* | MSK model *DoF \| Dim* | FES model | Cost function | Constraints | TRL |
|---|---|---|---|---|---|---|---|---|---|---|
| | | | | | | | | *AND FES activation* | | |
| Davoodi et al. (1999) [48] | *in silico* | N/A | Sit-to-stand | Hip flexors-extensors Knee flexors-extensors | Freq: N/A Width*: N/I Amp: N/A | 2 \| 2 | Durfee and DiLorenzo (1990) | *Min joint angle error OR min arm forces OR min FES activation* | *Task duration bounds* | 3 |
| Bao et al. (2020) [16] | *in vivo* | 2 AB 1 SCI | Sit-to-stand | Quadriceps | Freq: 35 Hz Width: 400 µs Amp*: 0-1 (-) | 2 \| 2 | First order dynamic AND Riener (1996) | *Min FES torque AND muscle fatigue AND device torque* | *Dynamics AND FES bounds AND Disturbance bounds* | 5 |
| Molazadeh et al. (2021) [49] | *in vivo* | 4 AB | Sit-to-stand | Quadriceps | Freq: 35 Hz Width: 400 µs Amp*: 0-1 (-) | 2 \| 2 | Veltink (1992) AND Riener (1996) | *Min FES torque AND device torque* | *Dynamics AND FES bounds AND Assistive device bounds* | 4 |
| Dosen and Popovic (2009) [50] | *in silico* | N/A | Walking | Ankle flexors-extensors Knee flexors-extensors | Freq: N/A Width: N/A Amp*: 0-1 (-) | 3 \| 2 | Veltink (1992) | *Min joint angle error AND min FES activation* | *Dynamics AND FES bounds* | 3 |
| Sharma and Stein (2011) [51] | *in silico* | N/A | Walking | Hip flexors-extensors Knee flexors-extensors | Freq: N/A Width*: N/I Amp: N/A | 2 \| 2 | First order muscle dynamics | *Min muscle activation AND walker reaction force AND ground collision* | *Dynamics AND Initial state AND Terminal state AND Kinematic bounds* | 3 |
| Sharma et al. (2014) [52] | *in silico* | N/A | Walking | Hip flexors-extensors Knee flexors-extensors | Freq: N/A Width*: 0-2000 (-) Amp: N/A | 3 \| 2 | First order muscle dynamics | *Min arm reaction force profile AND FES activation* | *Dynamics AND Initial state AND Terminal state AND Kinematic bounds AND Kinetic bounds AND FES bounds AND Task duration bounds* | 3 |
| Sauder et al. (2019) [53] | *in silico* | N/A | Fast treadmill walking | Medial gastrocnemius Semimembranosus Soleus | Freq*: 0-100% (gait cycle) and muscle | 37 \| 3 | Sauder 2019 | *Min joint jerk AND force asymmetry* | *Dynamics AND Terminal state AND Kinematic constraints* | 3 |

| Studies | Study design | Partici-pants | Lower limb task | Muscle stimulated | Reported stimulation settings* | MSK model<br>*DoF \| Dim* | FES model | Cost function | Constraints | TRL |
|---|---|---|---|---|---|---|---|---|---|---|
| | | | | Tibialis anterior | on-set/off-set<br>Width: N/A<br>Amp*: 0-0.7 (-) | | | | *AND Cyclicity constraints* | |
| Alibeji et al. (2018) [54] | *in vivo* | 1 AB<br>1 SCI | Walking | Hamstring<br>Quadriceps | Freq: 35 Hz<br>Width: 400 µs<br>Amp*: Min/max threshold | 4 \| 2 | Schauer (2005) AND Riener (1996) | *Min joint angle error AND min activation states AND synergy AND regularization* | *Dynamics AND Terminal state AND Kinematic bounds AND Muscle activation bounds AND Disturbance bounds* | 5 |
| Hodgins and Freeman (2023) [55] | *in vivo* | 3 AB | Walking | Tibialis anterior | Freq: 40 Hz<br>Width*: 0-300 µs<br>Amp: N/A | 1 \| 2 | Hammerstein model | *Min joint angle error AND min muscle fatigue AND compressor pressure* | *Dynamics AND FES bounds AND Assistive device bounds* | 4 |
| Li et al. (2018) [56] | *in vivo* | 3 AB<br>5 SCI | Walking | Lateral gastrocnemius<br>Tibialis anterior | Freq: 30-40 Hz<br>Width*: 0-450 µs<br>Amp: 25-35 mA | None | Zhang (2011) | *Min muscle activation error AND min aggressive control changes* | *N/I* | 5 |
| Nekoukar and Erfanian (2013) [57] | *in vivo* | 1 AB<br>2 SCI | Walking | Gastrocnemius<br>Gluteus maximus-minimus<br>Iliacus<br>Peroneal nerve<br>Quadriceps<br>Soleus | Freq: 25 Hz<br>Width*: 0-700 µs<br>Amp: Constant value | 10 \| 2 | Veltink (1992) | *Min joint angle error AND min muscle activation AND walker handle reaction force* | *Dynamics* | 5 |
| Gföhler et al. (2004) [58] | *in silico* | N/A | Cycling | Biceps femoris<br>Gluteus maximus<br>Iliacus<br>Psoas<br>Rectus femoris<br>Semimembranosus<br>Semitendinosus<br>Vastii | Freq*: 25-45 Hz<br>Width: N/A<br>Amp*: Min/max threshold | 3 \| 2 | Durfee and Palmer (1994) | *Max mean pedal torque* | *Kinematic bounds AND FES bounds* | 3 |

***Notes**: N/I - Not Informed by authors whereas it could be relevant, N/A - Not Applicable due to the nature of optimal control, AB – Able-bodied, SCI – Spinal cord injury, MSK - musculoskeletal, DoF - degree of freedom, dim - dimension 2D/3D, Min – Minimize, Freq – Frequency, Width – Pulse width, Amp – Pulse amplitude, (-) – normed, MRC - Medical Research Council Scale.*

*Table 2: Summary of studies* ***about upper limb movements*** *information (* optimized FES parameter).*

| Studies | Study design | Particip-ants | Upper limb task | Muscle stimulated | Reported stimulation settings* | MSK model<br>*DoF \| Dim* | FES model | Cost function | Constraints | TRL |
|---|---|---|---|---|---|---|---|---|---|---|
| Dunkelberger et al. (2022) [59] | *in silico* | N/A | Elbow flexion-extension | Biceps brachii<br>Triceps brachii | Freq: N/I<br>Width*: Min/max threshold<br>Amp: N/I | 1 \| 2 | Durfee and MacLean (1989) | *Min joint angle error*<br>*AND min FES activation*<br>*AND device torque*<br>*AND device torque derivative* | *Dynamics*<br>*AND Initial state*<br>*AND Kinetic bounds*<br>*AND Task duration bounds* | 3 |
| Kavianirad et al. (2023) [60] | *in vivo* | 5 AB | Elbow flexion-extension | Biceps brachii<br>Brachialis<br>Triceps brachii | Freq: 25 Hz<br>Width: N/I<br>Amp*: Min/max threshold | 1 \| 2 | Chen (2014) | *Min muscle activation error* | *Dynamics*<br>*AND FES bounds* | 4 |
| Wolf et al. (2023) [25] | *in silico* | N/A | Reaching | Biceps brachii<br>Brachialis<br>Deltoids<br>Infraspinatus<br>Latissimus dorsi<br>Lower and upper pectoralis<br>Rhomboids<br>Serratus anterior<br>Supraspinatus<br>Triceps brachii | Freq: 13 Hz<br>Width*: setting per muscle<br>Amp: setting per muscle | 4 \| 3 | Semiparametric Gaussian process regression model | *Min joint angle error*<br>*AND min muscle activation* | *Dynamics*<br>*AND Initial state*<br>*AND Terminal state*<br>*AND Kinematic bounds*<br>*AND Muscle activation bounds* | 3 |
| Dunkelberger et al. (2023) [61] | *in vivo* | 9 AB | Reaching<br>Grasping | Biceps brachii<br>Extensor carpi radialis longis<br>Extensor-flexor carpi ulnaris<br>Triceps brachii | Freq: N/I<br>Width*: Min/max threshold<br>Amp: N/I | 2 \| 2 | Durfee and MacLean (1989) | *Min joint angle error*<br>*AND min device torque derivative*<br>*AND FES activation*<br>*AND FES activation derivative*<br>*AND device control* | *Dynamics*<br>*AND FES bounds*<br>*AND Assistive device bounds* | 4 |
| Westerveld et al. (2012) [62] | *in vivo* | 2 AB | Reaching<br>Grasping<br>Releasing | Abductor pollicis longus<br>Finger extensors-flexors<br>Flexor pollicis brevis<br>Opponens pollicis | Freq: 30 Hz<br>Width: 80 µs<br>Amp*: Min/max threshold | ARX model | Second order linear dynamic polynomial model | *Min joint angle error* | *FES bounds*<br>*AND Task duration bounds* | 4 |
| Westerveld et al. (2014) [63] | *in vivo* | 2 AB<br>2 Stroke | Reaching<br>Grasping<br>Releasing | Abductor pollicis longus<br>Extensor digitorum communis<br>Flexor digitorum superficialis<br>Flexor pollicis brevis<br>Opponens pollicis | Freq: 30 Hz<br>Width: 80 µs<br>Amp*: Min/max threshold | ARX model | Second order linear dynamic polynomial model | *Min joint angle error* | *FES bounds* | 5 |
| Arrofiqi et al. (2021a) [64] | *in silico* | N/A | Palmar flexion | Flexor carpi radialis | Freq: N/I<br>Width: N/I<br>Amp*: N/I | 1 \| 2 | Average response to FES model | *Min joint angle error AND min FES activation difference* | *N/I* | 3 |
| Arrofiqi et al. (2021b) [65] | *in silico* | N/A | Palmar flexion | Flexor carpi radialis | Freq: N/I<br>Width: N/I<br>Amp*: 0-1 (-) | ARX model | Average response to FES model | *Min joint angle error AND min FES activation difference* | *FES bounds* | 3 |
| Arrofiqi et al. (2023) [66] | *in silico* | N/A | Palmar flexion | Flexor carpi radialis | Freq: N/I<br>Width: N/I<br>Amp*: 0-1 (-) | 1 \| 2 | Arrofiqi (2021) | *Min joint angle error AND min FES activation difference* | *FES bounds* | 3 |

***Notes:*** *N/I - Not Informed by authors whereas it could be relevant, N/A - Not Applicable due to the nature of optimal control, AB – Able-bodied, MSK - musculoskeletal, DoF - degree of freedom, dim - dimension 2D/3D, ARX - Autoregressive with Extra Input, Min – Minimize, Freq – Frequency, Width – Pulse width, Amp – Pulse amplitude, (-) - normed.*

### 3.1 FES modeling

The most frequently used FES models were the model of Hunt et al. (1998) [67] (13%, n = 6), the combined use of the Veltink et al. (1992) [68] and Riener et al. (1996) [69] models (11%, n = 5 studies), and the model of Ding et al. (2003) [15] (11%, n = 5) (Fig. 2). Most models were adopted by a single research team, except Veltink et al. (1992)'s model (3 teams) and Ding et al. (2003) (2 teams). These models can be grouped into three categories: 1) recruitment-activation models (59%, n = 26 studies), often a linear muscle recruitment equation combined with a nonlinear muscle activation model, 2) data-driven nonlinear models (27%, n = 12), designed to map muscle activations to obtain force production and 3) force-fatigue phenomenological models (17%, n = 6), capable of grasping force decay and recovery. Only 10 studies (22%) provided a rationale for their chosen FES model (available in supplementary File 2).

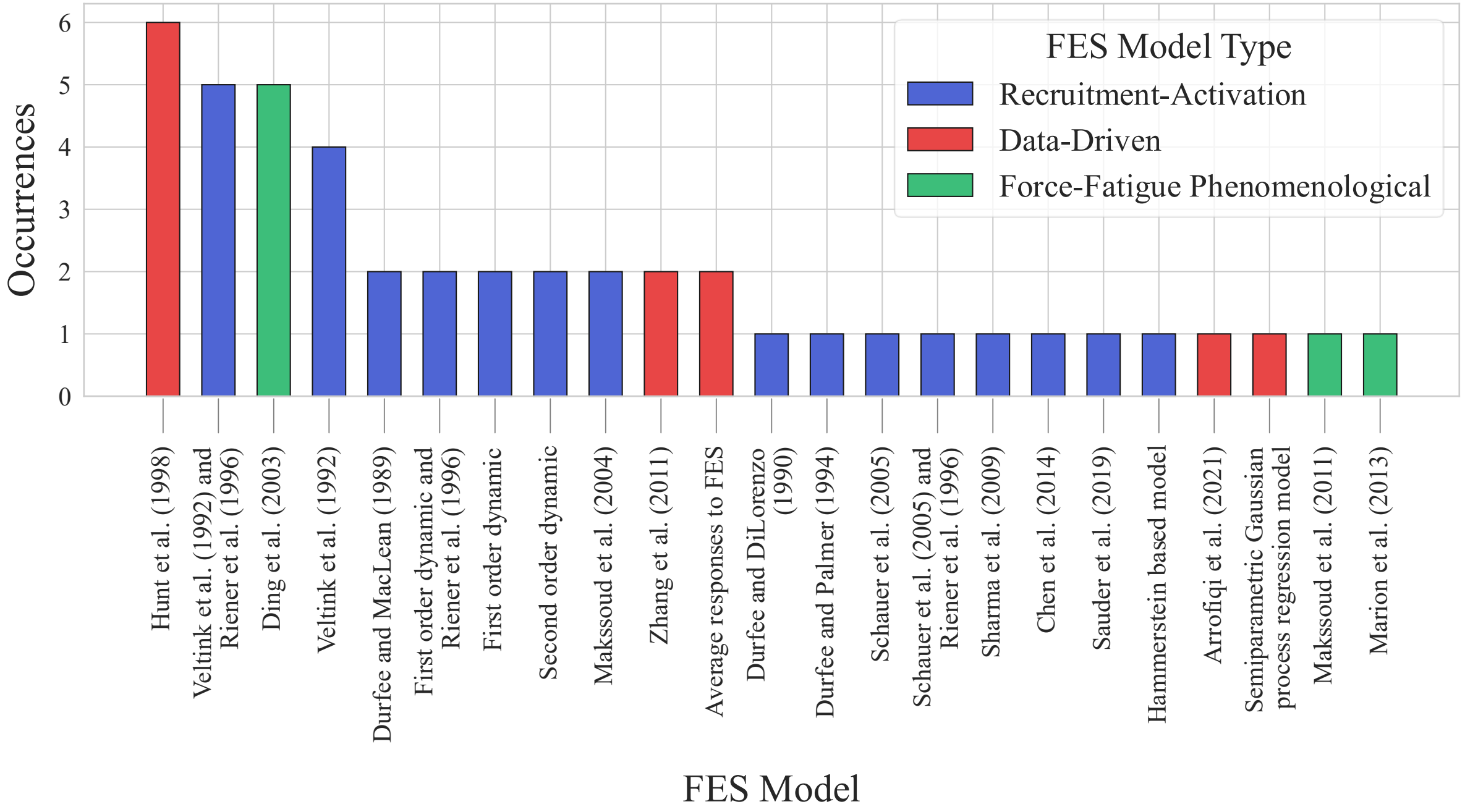

*Figure 2: Occurrences of FES model used*

Using the proposed logarithmic scoring system to assess FES model complexity, the observed models took three values: 0 (n = 31), 1.17 (n = 8) and 1.79 (n = 5), termed as simple, medium, and complex. Early studies predominantly relied on simple FES models, likely due to hardware constraints (Fig. 3), medium and complex models combined with optimal control only emerged over the last ten years. To date, no study has implemented a complex FES model with a detailed musculoskeletal model or complex optimal control program. Real-time optimization could not be achieved whenever either the FES or MSK models were too complex.

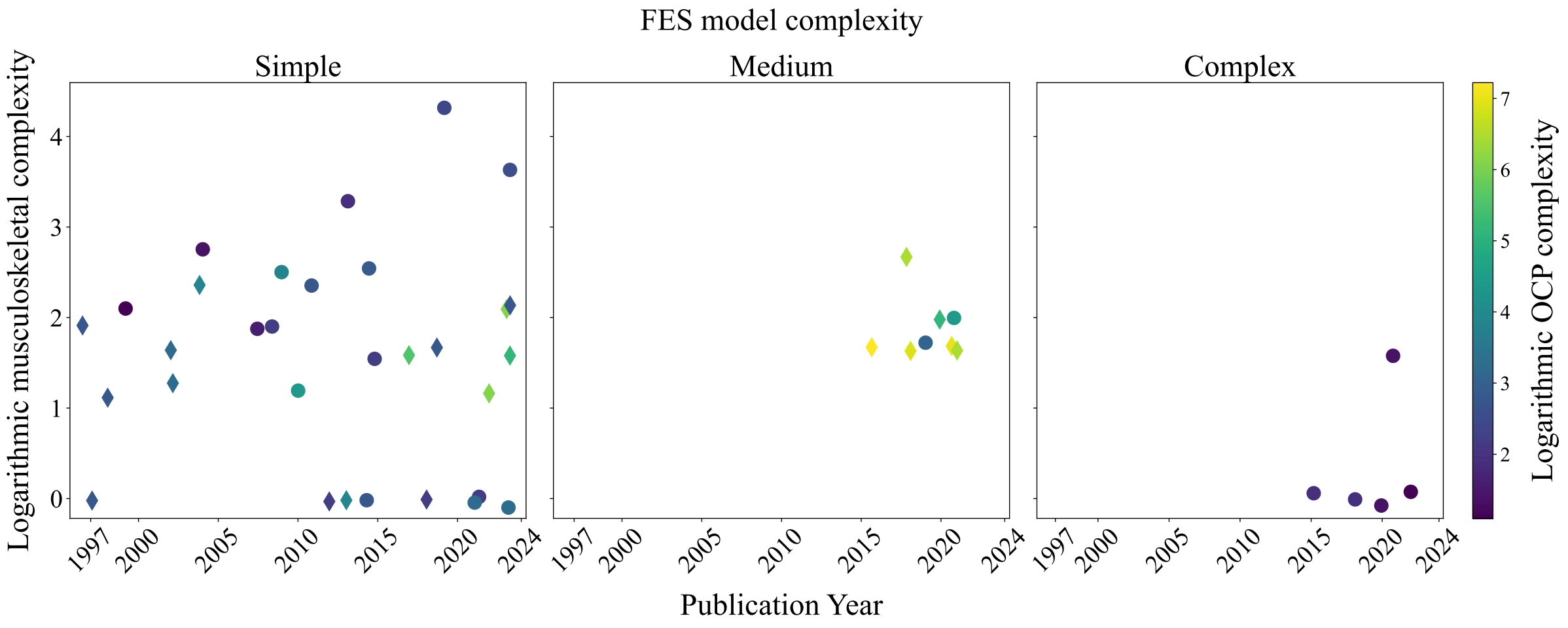

*Figure 3: Logarithmic score representation encompassing musculoskeletal model complexity (y-axis), optimal control problem (OCP) complexity (color gradient) and FES model complexity (subplot) per publication year. With ♦ for real-time and ● for offline or non-reported computation time.*

### 3.2 Optimal control

Across studies, the definition and implementation of OCP varied. Concerning the control variables, the most frequently used were pulse intensity (48%, n = 21 studies) and pulse width (45%, n = 20) followed by onset-offset timing (9%, n = 4), frequency (7%, n = 3) and dynamic muscle targeting (2%, n = 1).

Concerning the cost functions, they can be grouped into two categories: 1) data tracking and 2) minimization (Fig. 2), which can be subdivided according to involved variables: 1) tracking of joint angle (25%, n = 26 studies) and muscle activation (9%, n = 9), and 2) minimization of muscle activation (7%, n = 7), FES parameters (12%, n = 13) and assistive device activations (6%, n = 6).

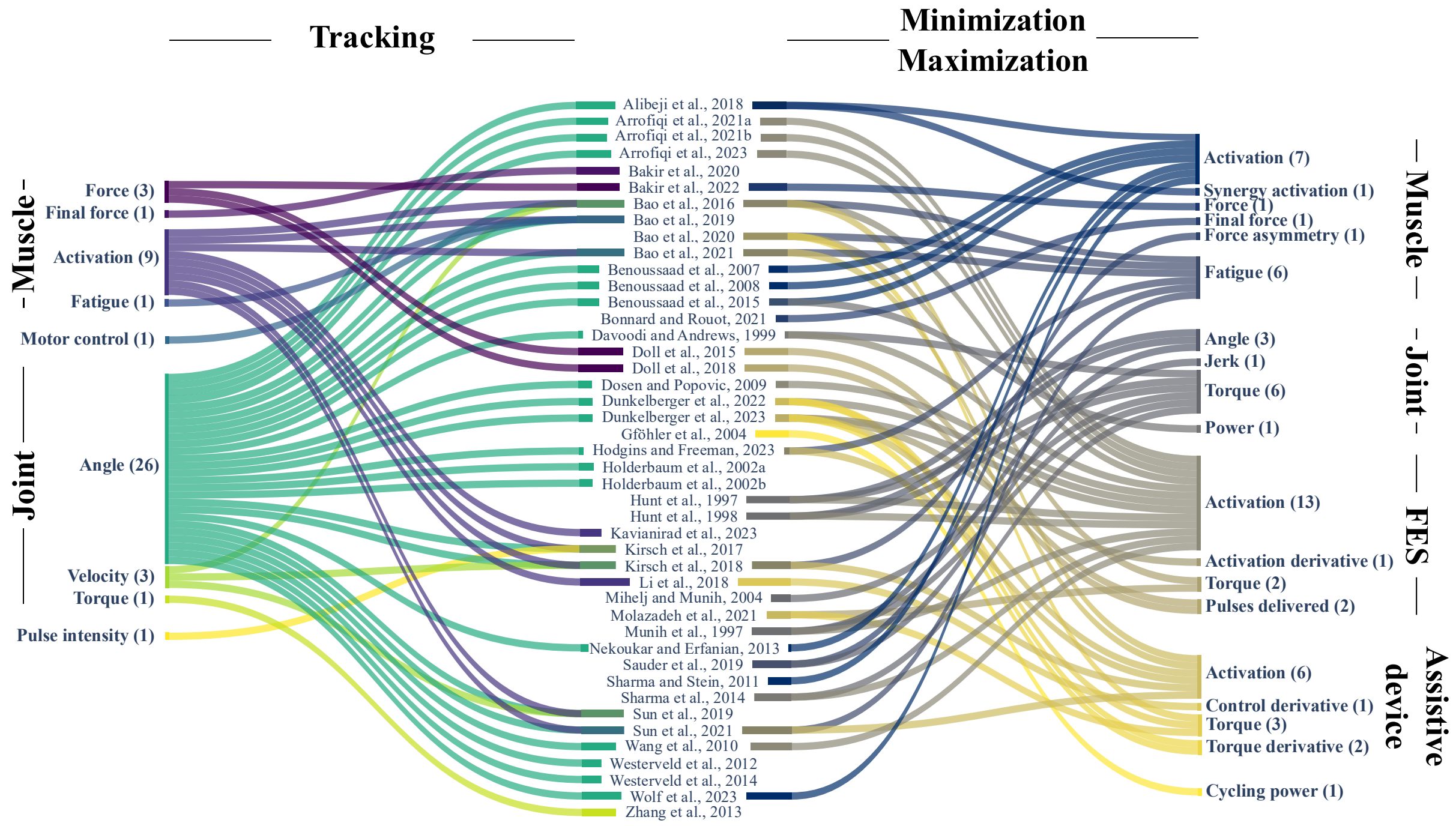

*Figure 4: Cost functions used in the FES-based optimal control problems sorted by types (tracking on the left, minimization/maximization on the right) and then subcategories, namely, muscle, joint, FES, assistive device.*

A set of constraints was associated to these cost functions and can be grouped into three categories: **1)** satisfaction of the dynamic equations inherent to OCPs (59%, n = 26 studies explicitly reported information about this constraints), **2)** boundaries for FES control (61%, n = 27), kinematics (25%, n = 11), assistive device control (14%, n = 6), task duration (14%, n = 6), muscle activation (11%, n = 5) and disturbance (9%, n = 4), and **3)** initial (23%, n = 10) and terminal (20%, n = 9) constraints set to satisfy the task.

Control strategies were most often implemented in a closed-loop scheme (61%, n = 27 studies) with gradient-based solvers (50%, n = 22). Meanwhile, 20% (n = 9) of studies did not specify the employed solver. Numerical integration methods (e.g., Runge-Kutta) were reported in only 23% (n = 10 studies), without any method prevailing. The discretization intervals averaged 20 ms (range: 1-100 ms), though 57% (n = 25) of studies omitted this detail. Real-time implementations were possible in 48% (n = 21) of studies, offline in 18% (n = 8), and 34% (n = 15) did not report the information. Only 16% (n = 7) of studies provided explicit computation times. Close to a third of optimal control-driven FES (30%, n = 13 studies) used hybrid FES / external assistance (e.g., exoskeletons, prostheses) control scheme. MATLAB was the predominant environment for optimal control implementation (64%, n = 28 studies). Details for each study are available in supplementary File 2.

### 3.3 Muscle fatigue

According to the proposed levels to assess muscle fatigue inclusion, results indicate 16% studies (n = 7) at level 0, 16% (n = 7) at level 1, 2% (n = 1) at level 2, 16% (n = 7) at level 3, 11% (n = 5) at level 4, 11% (n = 5) at level 5, 11% (n = 5) at level 6, 14% (n = 6) at level 7, and 2% (n = 1) at level 8. The details for each question are available in supplementary Fig 5.

Muscle fatigue was directly measured in four studies (9%). Experimental comparisons were: *i)* a constant frequency pulse train vs. an optimized pulse train during an isometric task [28,29]; *ii)* Proportional integral derivative vs. nonlinear model predictive control for an isometric task [37]; and *iii)* FES plus motor assistance vs. FES alone for a knee extension [38]. In each case, muscle fatigue was mitigated when using optimal control compared to a suboptimal method.

### 3.4 Optimal control-driven FES's technology readiness level

Half of the studies were conducted purely *in silico*, corresponding to TRL 1–3. These include works focused on defining OCPs (TRL 1, 2%, n = 1 study) [17], generating feasible stimulation patterns associating them with biomechanical models (TRL 2-3, 48%, n = 21) [32]. Above TRL 3, only 23% of studies (n = 10) reached the TRL 4 level by conducting proof of concept experiments with healthy individuals in controlled laboratory environments, and 27% of studies (n = 12) reached a TRL 5 level by conducting validation involving functional tasks with neurological impaired subjects while remaining in controlled laboratory environments. None of the included studies involved longitudinal protocols, real-world clinical integration, or external replication by independent teams, i.e. key elements required to reach TRL 6 and above.

## 4. Discussion

This scoping review examined the application of numerical optimal control methods to FES, highlighting their potential to personalize stimulation protocols and reduce muscle fatigue. The main findings are *(i)* a high degree of variability in both FES models and *(ii)* in the design and implementation of OCPs, with *(iii)* limited efforts to assess and mitigate muscle fatigue. Overall, *(iv)* the TRL remains low with limited clinical relevance. Finally, *(v)* FES model identification, real time optimization and commercial hardware customization were identified as barriers limiting the development of the practice and the related clinical translation.

These findings should be interpreted with caution, considering three main limitations. First, in accordance with the PRISMA-ScR guidance, we did not perform a formal risk-of-bias or quality assessment, which may temper the strength of our conclusions. Second, by excluding any study that did not employ an explicit FES model (to quantify muscle activation, force, or torque [21]) to maintain a tight focus on FES-specific optimization, relevant work in a wider neuromuscular or hybrid-control contexts might have been overlooked. Third, we made some classification choices to provide a broad overview of the field, but we may have oversimplified certain features, particularly in the design and formulation of the cost functions and constraints used across studies.

The following sections detail the key findings with a focus on the challenges in FES modeling, in optimal control for FES, and in preparing such applications for clinical translation. Recommendations and perspectives are given for each section (Table 3).

*Table 3: Recommendations and perspectives summary*

| Number | Recommendations |
|---|---|
| **R1** | Justifying the FES model choice. |
| **R2** | Clearly and fully describing the optimization problem. |
| **R3** | Providing all information concerning the numerical framework. |
| **R4** | Using open-source prototyping environments to conduct research. |
| **R5** | Using shared benchmarks and standardized protocol for cross study comparison. |
| **R6** | Sharing data, code and musculoskeletal models to promote open science and reproducibility. |
| **Number** | **Perspectives** |
| **P1** | Conducting comparative studies on FES models. |
| **P2** | Enhancing FES models computational efficiency and identification. |
| **P3** | Coupling FES models with musculoskeletal models in the framework development. |
| **P4** | Developing and validating FES fatigue models encompassing multiple control variables capable of dissociating fatigue from force or activation. |
| **P5** | Improving control robustness to variability, model uncertainty, and patient voluntary contraction. |
| **P6** | Identifying the most effective cost function for a given task and rehabilitation purpose. |
| **P7** | Conducting a review on reinforcement learning for FES. |
| **P8** | Improving FES models identification procedure. |
| **P9** | Performing longitudinal investigations to evaluate optimal control-driven FES benefits. |
| **P10** | Developing commercial stimulators allowing a personalized and deep control of FES for rehabilitation. |
| **P11** | Enhancing hybrid systems to improve rehabilitation. |

### 4.1 Challenges in FES modeling

The purpose of FES models is to predict muscle responses (e.g., force, fatigue, torque) for specific stimulation inputs (e.g., frequency, pulse width, intensity). Like musculoskeletal model [70], the FES model selection can dramatically influence both the motor task execution and the study's conclusions. This section aims to highlight current trends and proposes a set of recommendations and research perspectives concerning FES modeling.

#### *4.1.1 Current trends in FES modeling*

Across the 44 included studies, 23 distinct FES models were used (Fig. 3), highlighting a significant variability and a lack of standardization in the field. However, by identifying three categories of models, we can define current trends. Recruitment-activation models, such as in Veltink et al. (1992) [68] or the first and second order models are especially suited for real-time close-loop control, where fast model identification is more critical than high fidelity or muscle fatigue estimation [71]. In terms of complexity, these models were classified as *Simple or Medium* complexity, according to their formulations and required parameters. Data-driven nonlinear models, often built with the Hammerstein structure (e.g., Hunt et al. (1998) [67]), are particularly effective for real-time, high-precision and closed-loop FES control. All FES models in this category were classified as *Simple*. Force-fatigue phenomenological models are typically used for open-loop control where real-time optimization and identification procedures are not a concern, enabling the estimation of force decay, and helping optimize stimulation to mitigate fatigue [29]. Complexity levels for this category ranged from *Medium* to *Complex*. Notably, none of the studies using *Complex* models achieved real-time implementation, further supporting their role in offline or open-loop systems. Among them, the Ding et al. (2003) model [15] remains the benchmark for capturing stimulation-dependent muscle properties during long tasks.

#### *4.1.2 Validation of used FES model*

Like musculoskeletal models, FES model validation remains a major challenge, as it requires a large amount of data and is difficult to implement. Data-driven models are directly trained with subject- or population-specific responses to FES, meaning that the validation process is effectively built into model development. In contrast, recruitment-activation models require validation. Indeed, while the recruitment phase is typically linear between contraction threshold and muscle saturation, the nonlinear activation component requires validation. The model of Veltink et al. (1992) [68], frequently cited in this review, was originally validated on isometric contractions in cat soleus muscle. Despite this, it has been widely used in human *in vivo* studies, often without validation for the task [57], muscle group [38], or population [16]. This reflects a broader issue: many FES models may not be validated for their intended research applications. Finally, force-fatigue phenomenological models require full-model validation, especially for predicting long-term stimulation effects. To date, only two studies applied such models *in vivo* [28], both within the validated scope (healthy human subjects, quadriceps muscle). This section offers only a limited overview, as the topic warrants a dedicated review [72].

#### *4.1.3 Recommendations and research perspectives for FES modeling*

Only 20% of the reviewed studies provided a rationale for their FES model selection. To promote reproducibility and informed decision-making, we recommend that authors explicitly justify their model choice (recommendation R1, Table 3) based on the study context [73], e.g. selecting force-fatigue models for fatigue-related investigations, or recruitment-activation models for real-time closed-loop control. There is also a clear need for comparative studies across FES model categories (perspective P1), similarly to those conducted for musculoskeletal models [70], as in the study proposed by Frey Law and Shields [74]. Such analyses would clarify the trade-offs between complexity, computational efficiency, and potential effects on muscle fatigue mitigation. Although medium and complex models are less common than simple models, they are often applied beyond their validated scope, and their added value remains uncertain. Broader benchmarking could help determine whether increased model complexity truly improves fatigue mitigation or motor performance. The limited clinical adoption of FES models is partly due to real-time constraints and time-consuming implementation of protocols [75,76] Simplifying models according to the parsimony principle [77]and embedding online parameter identification into therapy sessions (P2) could help overcome these barriers [26,27]. Combining FES models with musculoskeletal simulations (P3) and developing multi-variable frameworks that dissociate force and fatigue (P4) could further improve control strategies and support more effective rehabilitation outcomes. A key feature could be to partially dissociate muscle force or activation from muscle fatigue, as different combinations of FES settings can lead to a similar torque production with different fatigue states [78,79].

## 4.2 Challenges in optimal control for FES

Optimal control offers an efficient numerical framework that integrates performance goals (Fig. 4, using tracking or minimization/maximization cost functions) and safety constraints (e.g., FES control and joint ranges of motion boundaries) into a single optimization problem [25,32,34,39,54]. It accommodates various models (Fig. 3), motor tasks (Tables 1 and 2), and assistive technologies (e.g., exoskeletons [33]), enabling personalized and physiologically coherent FES control. This section outlines current trends and provides recommendations for advancing optimal control in FES.

### *4.2.1 Current trends in the optimization problem formulation*

Control variables have a significant influence, particularly in the context of FES, as they directly affect muscle behavior and clinical outcomes [9,79–81]. Five control variables were identified. Pulse intensity [42] is the most used but has limited effectiveness in reducing fatigue [80] and is associated with pain and discomfort [79,82], making it less suitable for rehabilitation, despite its ease of normalization. Pulse width [17] is also widely used and more effective for fatigue management [79]. Frequency [28] and stimulation on/off timing [26] were explored in only a few studies, mainly due to the limited availability of models supporting these variables, often restricted to force-fatigue phenomenological models. Dynamic muscle targeting, i.e. the ability to vary which muscles are stimulated during a task, was explored in only one study [53] but holds promise for personalized stimulation protocols. Few studies have explored multi-variable control strategies [57]. Only one simulation study combined both pulse timing and width [27]. Most others focused on a single variable [29,55,63] sometimes alongside external devices [34,54]. While pulse width and pulse intensity are the two dominant approaches in FES optimization, multiple-variable FES control remains underexplored.

Cost functions play a central role in shaping how OCP prioritizes task performance and physiological constraints in FES. While kinematic costs (e.g., joint angle tracking) appeared to be widely used and effective for simple, single-joint movements, they may limit the ability to explore alternative motor strategies [83] or take advantage of muscle redundancy to delay fatigue [84]. In contrast, muscle-based cost functions (e.g., activation, force, or fatigue) offer a more direct control over physiological responses to reduce overstimulation [42]. Combining both types in multi-objective formulations allows optimal control to balance competing goals, such as ensuring task accuracy while minimizing fatigue [65] or the reliance on assistive devices [32]. This integrative approach is key to advancing FES strategies toward more adaptable and patient-specific applications.

Constraints and kinematic, FES, and assistive devices bounds have a central role in OCP. They are commonly used to ensure patient safety and help to adjust the stimulation to the individual's physiological characteristics while complying with hardware capabilities [27–29]. In addition, soft constraints, implemented as cost terms, can support real-time applications by easing convergence while being dynamically correct [33]. No studies have employed adaptive or time-evolving boundaries for longer tasks, to further improve realism by accounting for muscle fatigue or stimulation familiarization.

#### *4.2.2 Current trends in the numerical framework employed*

Early OCP implementations generated open-loop stimulation patterns offline [58]. By contrast, recent works increasingly adopt closed-loop, model-predictive control (MPC), which updates stimulation in real-time from force or kinematic feedback and thus better handles fatigue and other time-varying dynamics [35,60,61,66]. Most included studies that detail their numerical integration scheme favor explicit Runge–Kutta methods, valued for coding simplicity. While collocation, which has a better convergence rate, is often more computationally efficient and provide overall better results [85], remains under-used [86].

Concerning the discretization, it was typically reported on the order of 20 ms, a compromise between accuracy and time-solving iterations. It must be noticed that a discretization step should be at least as precise as the stimulation frequency [17], but models incorporating stimulation instants may require even smaller steps to capture fast state transitions between pulses [15]. In some cases, adaptive discretization has also been employed to meet specific optimization demands requiring a more complex OCP formulation [36].

Real-time implementation is a concern for clinical applications as it is essential to achieve accurate control in practice [75], but only a few studies reported actual time performance. Optimization time depends on problem complexity (Fig. 3) and solver iteration calculation rate. Solver choice significantly impacts performance, while general-purpose solvers (e.g., IPOPT) are widely used, they may not be optimized for fast execution. Tailored solvers such as real-time QP/NLP solvers or code-generated frameworks (e.g., ACADOS) offer faster convergence and better integration with embedded systems [87,88]. However, few studies explicitly report solver configuration or computational benchmarks, limiting reproducibility and clinical readiness.

#### *4.2.3 Recommendations and research perspectives for optimal control*

The impact of optimal control in FES remains limited by current models and hardware. Broader integration of stimulation parameters in models (P4) and improving hardware flexibility would increase clinical relevance. Clear reporting of OCP definitions, including control variables, cost functions, constraints (R2), and numerical details like integration methods, solvers, computation times, hardware specification (R3), is essential for reproducibility. The work of Wolf et al. (2023) [25] provides a valuable example. While MATLAB remains the dominant prototyping environment, its licensing constraints may hinder clinical deployment. Future work should consider open-source alternatives better suited for translation (R4). Improving robustness to variability, model uncertainty, and patient voluntary contraction will be critical for reliable real-world use (P5) [33] Although closed-loop control is preferred clinically, open-loop methods remain useful for testing [25] and refining [53,62] cost functions. To showcase the true potential of optimal control, comparative studies of optimal strategies using a robust measurement metric are required. These studies must both demonstrate clear improvements in muscle fatigue reduction over classical stimulation and identify the most effective configurations (P6). Finally, reinforcement learning, that was out of the scope of this review, warrants a dedicated review given its rapid growth and potential (P7). Optimal control has shown the capability to evolve to the needs of rehabilitation, making this control a promising method for the future evolution of FES.

### 4.3 Challenges for clinical translation

#### *4.3.1 How close is optimal control-driven FES to clinical use?*

Despite promising developments, optimal control-driven FES has remained at an early stage of clinical readiness. Based on TRL classification, half of studies remain at TRL 1–3. Notably, none demonstrated longitudinal use, integration in clinical workflows, or independent replication, which are milestones toward clinical adoption. Optimal control-driven FES is already supported by existing scientific evidence, with most models grounded in established theory and validated *in silico* (TRL 1–2). Interestingly, a transition toward *in vivo* research has started. However, validation remains narrow in scope.

Most experimental protocols involved young, able-bodied individuals, with a few studies recruiting clinical populations. Among pathological cohorts, SCI was overrepresented, while stroke, despite accounting for a significant proportion of neurorehabilitation cases, remains underexplored. A strong sex imbalance was also observed. Although the male-to-female ratio reflects the epidemiology of SCI (3.8:1) [89], it remains disproportionately high and limits the generalizability of findings. Given known sex-related differences in neuromuscular response and recovery [90,91], future studies should aim for more balanced recruitment.

From a motor task perspective, most studies targeted lower limb movements, which aligns with the higher prevalence of paraplegia in SCI [89]. The quadriceps was the most stimulated muscle (36% of lower-limb studies), reflecting a preference for accessible, superficial targets. This focus also underscores interest in engaging a major contributor to walking and cycling, where muscle fatigue reduction is crucial for rehabilitation [92]. Deeper or more functionally complex muscle groups remain underrepresented. Likewise, nearly all studies used surface electrodes; only one explored implanted stimulation [36].

Altogether, these findings position optimal control-driven FES at an early TRL 5 out of 9 stages, with validation largely limited to controlled laboratory settings, and minimal clinical integration. Significant gaps remain in population diversity, real-world applicability, and experimental robustness, i.e. critical areas to address for advancing toward clinical translation.

#### *4.3.2 Recommendations and research perspectives in preparing for clinical translation*

To accelerate the clinical adoption of optimal control-driven FES, advances are required in personalization, validation, and integration. A central priority is the development of patient-specific rehabilitation strategies (P8), notably by automating FES model parameter identification or embedding it directly into clinical workflows. This would reduce patient burden and support more adaptable, real-world implementations. Despite growing interest, no study has yet assessed the long-term effects of optimal control-driven FES in rehabilitation. Longitudinal investigations with larger cohorts are needed to evaluate sustained benefits such as fatigue reduction or improved motor recovery, which are essential to reach TRL 6 (P9).

Similarly, integration into clinically certified FES systems remains limited. Demonstrating compliance with safety and regulatory standards (e.g., ISO 14971, IEC 62304) is a prerequisite for clinical deployment (TRL 8) (P10). Currently, only three devices offer both programmable APIs and regulatory clearance: the Rehastim stimulator (Hasomed, Germany), used in five studies of this review, the MotiMove stimulator (3F-Fit Fabricando Faber, Serbia) with a clearance exclusive to Europe [93] and the g.Estim FES stimulator (g.tec medical engineering, Austria). Hybrid assistive systems, which combine FES with robotic supports or exoskeletons, offer a promising avenue for achieving functional gains in more demanding motor tasks (P11). Finally, establishing shared benchmarks, including standardized protocols (R5), datasets, or open-source models (R6), would support cross-study comparisons and facilitate the validation of early-stage work (TRL 1–3). Such efforts should increase transparency and reproducibility (R1–R6), as promoted by the International Functional Electrical Stimulation Society community.

## 5. Conclusion

This scoping review highlights broad variability in FES models and optimal control approaches, with a predominance of simple recruitment–activation schemes, single-joint, and tracking cost functions. Notably, strategies for muscle fatigue mitigation, rigorous model validation, and real-time closed loop implementation remain rare. Key obstacles to clinical translation include inadequate model identification, computational demands, hardware constraints, and solution robustness. To advance the field, future work should focus on (1) developing and validating FES models, (2) designing multi-objective optimization that balances movement accuracy and fatigue reduction, (3) creating standardized benchmarks and open-source frameworks, and (4) integrating these solutions into real world clinical workflows. Collaborative, transparent efforts will be essential to realize the rehabilitation potential of optimal control–driven FES.

**CRediT authorship contribution statement**

**Kevin Co:** Conceptualization, Formal analysis, Investigation, Methodology, Visualization, Writing – original draft, Writing – review and editing. **Mickaël Begon:** Conceptualization, Methodology, Supervision, Validation, Writing – review and editing. **François Bailly:** Validation, Writing – review and editing. **Florent Moissenet:** Conceptualization, Methodology, Supervision, Validation, Writing – review and editing.

**Ethics statement**

An approval by an ethics committee was not applicable for this study.

**Funding**

This research was supported by the Fonds de recherche du Québec – Nature et technologies (FRQNT).

**Declaration of competing interest**

The authors declare that they have no known competing financial interests or personal relationships that could have appeared to influence the work reported in this paper.

**Declaration of generative AI and AI-assisted technologies in the writing process**

During the preparation of this work the authors used ChatGPT in order to improve the manuscript clarity and readability. After using this tool/service, the authors reviewed and edited the content as needed and takes full responsibility for the content of the publication.

**Acknowledgement**

The authors would like to thank Prof. Christine Azevedo-Coste for her guidance on the design of the scoping review and her expertise in functional electrical stimulation. We also extend our gratitude to Mr. Denis Arvisais for his assistance with the scoping review search strategy and the retrieval of reviewed articles.

**Supplementary files**

1) *Literature search report*: *Literature Search Report.docx*
2) *Data extraction: scoping_review_for_article.xlsx*

## Supplementary figure

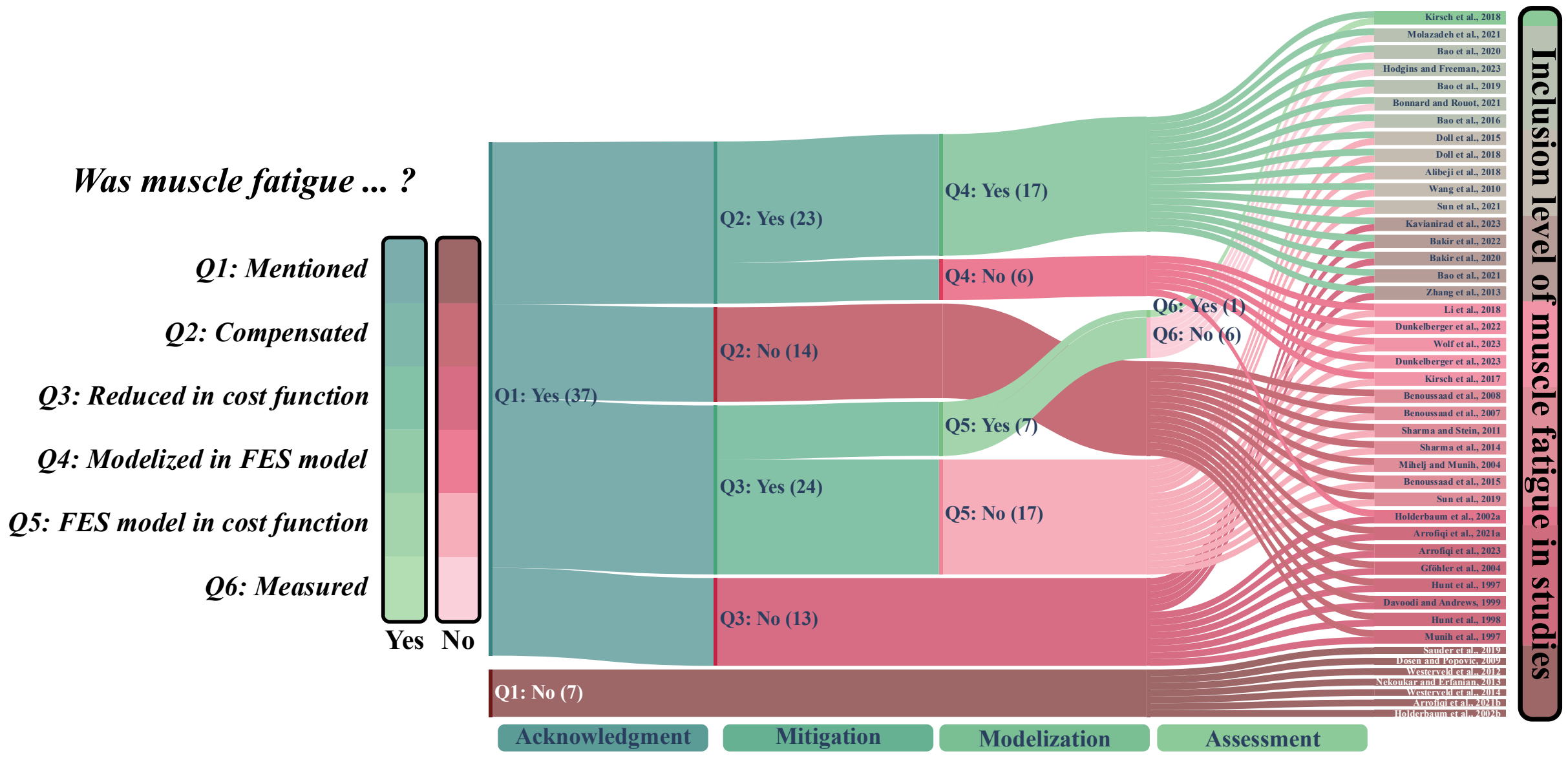

*Figure 5: Muscle fatigue inclusion for each article*

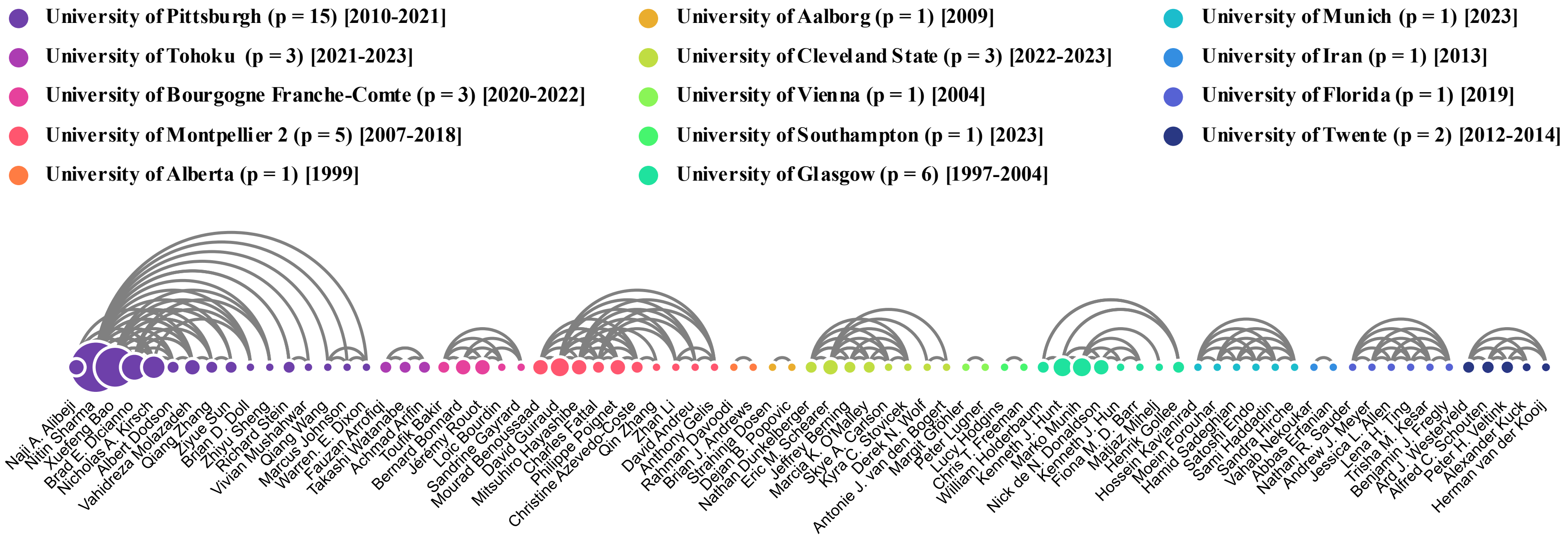

*Figure 6: Collaboration on OC-FES driven research, main contributor's university, p the number of publications, latest to newest publication date*

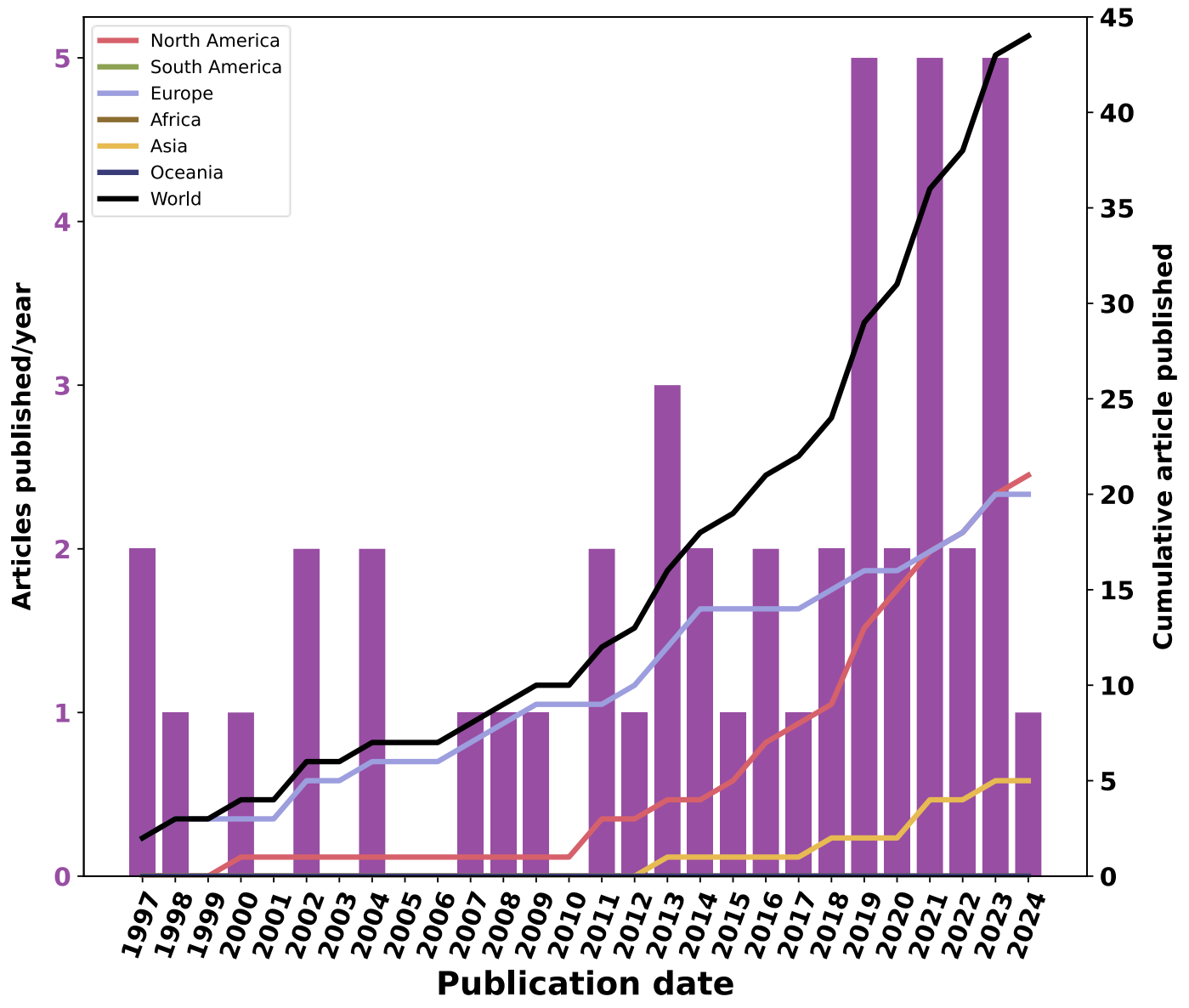

*Figure 7: Publications per year and regions*